\documentclass[epj]{svjour}
\usepackage{graphicx}
\usepackage{amsmath}
\begin{document}
\title{Physical properties of  voltage gated pores}
\author{L. Ram\'irez--Piscina\inst{1} \and J.M. Sancho\inst{2}}
\institute{Departament de F\'isica, Universitat Polit\`ecnica de Catalunya,
Avinguda Doctor Mara\~n\'on 44, E-08028 Barcelona, Spain \and Departament de F\'isica de la Mat\`eria Condensada, Universitat de Barcelona, Universitat de Barcelona Institute of Complex Systems (UBICS),
Mart\'i i Franqu\'es 1, 08028 Barcelona, Spain}

\date{\today}
\abstract{
Experiments on single ionic channels have contributed to a large extent to our current view on the function of cell membrane. In these experiments the main observables are the physical quantities: ionic concentration, membrane electrostatic potential and ionic fluxes, all of them presenting large fluctuations. The classical theory of Goldman--Hodking--Katz assumes that an open channel can be well described by a physical pore where ions follow statistical physics. Nevertheless real molecular channels are active pores with open and close dynamical states. 
{ By skipping the molecular complexity of real channels, here we present the internal structure and calibration of two active pore models.}
These models present a minimum set of degrees of freedom, specifically ion positions and  gate states, which follow Langevin equations constructed from an unique potential energy functional and by using standard rules of statistical physics.
Numerical simulations of both models are implemented and the results show that they have dynamical properties very close to those observed in experiments of Na and K molecular channels. 
In particular a significant effect of the external ion concentration on gating dynamics is predicted, which is consistent with  previous experimental observations. { This approach can be extended to other channel types with more specific phenomenology.}}

%
\maketitle

\section{Introduction}

Ionic transport across the cell membrane is a very common process in all cells.
In most situations the transport is done by very specific pores called molecular channels. These devices are embedded into lipid membranes and operate between different ionic concentrations at both sides, and subjected to the membrane voltage. 
Molecular channels are not passive  
pores but they have internal structures (gates) that present conformational states  such as open and close configurations.
When the channel is in the open state it allows the flux of charges driven by the ionic  density gradient and the membrane electrostatic potential. This flux can modify also this potential. Moreover the whole process is very selective: only a particular ion can cross a specific channel. 
In the close state no flux of ions is in principle allowed, except for some small leak. 
The transitions between these two conformational states are controlled by the membrane potential and thermal fluctuations. 

The synchronized dynamics of a large number of such channels of several types, coupled to the ionic concentrations at both sides of the cellular membrane, provides mechanisms for action potentials in neurons, cardiac cells, etc \cite{Hille,Hammond,phillips}.
Beyond the study of the conductivity properties of the membrane (i.e. resulting from a large number of channels), experiments performed on single channels  have provided a good deal of information on the gating dynamics of individual channels \cite{Hammond}.  Such experiments have shown that, in addition to the inherent randomness of the gate open-close dynamics, there are also very strong fluctuations in the charge flux.

The biochemical structure of a molecular channel is quite complex at the molecular level.
From the physical point of view the relevant observables of a channel are the ionic concentrations, the membrane electrostatic potential, and the ionic fluxes. Here 
we will address the dynamics of these observables by modeling the channel as a simple active pore, with a reduced set of observables fulfilling known physical laws and consistently incorporating thermal fluctuations.

Most theoretical modelings follow  the Hodking-Huxley approach \cite{H-H} for the dynamics of 
membrane permeability. 
Fluctuations have been incorporated 
by using either Langevin or master equations \cite{jung2001,schmid2001,ozer2009,groff2009markov,huang2011}.
In previous publications, 
a semi-microscopic Langevin modelization  was used to show the excitable properties of a single pore mimicking a Na channel in the presence of a leak of K ions \cite{Ramirez1}, and the periodic firing of a pair of Na, K--like active pores \cite{ramirez-periodic}. 
In this approach all microscopic relevant variables (a reduced set of degrees of freedom) follow stochastic (Langevin) dynamical equations according to very general principles of statistical physics.
{ 
The main aim is to 
use only the necessary physical mechanisms involved, and to obtain their effects by using standard physical laws applied in a consistent way. Additional biological complexity would indeed be necessary to explain more specific physiological features, but it is here stripped away in order to identify the essential elements for explaining the basic channel gating phenomenology observed in experiments. }

By using this approach we will study in detail the stochastic dynamics of two different pore models.
Although we will reduce the elements of the modeling to the minimum necessary to account for the behavior of single channels, we will show that the dynamics do present the behavior observed in Na and K channel experiments. Moreover the physical consistency of the model results to be a key element in such agreement.

The outline of this paper is as follows.
In the following section, we summarize the theoretical structure of the approach. In the next section  we present the calibration and dynamical properties of two pore models, and its comparison with the available experimental information.
Finally we end with some conclusions and perspectives. The specific technical details of the approach are presented in the Appendix.

\section{Gating pore models}

\subsection{Summary of the approach}

%

Our approach \cite{Ramirez1} follows the path opened by the Goldman--Hodking--Katz (GHK) equation \cite{Goldman} and the Hodking--Huxley theory \cite{H-H}. 
It
is formulated by constructing an energy functional modeling the interaction between all variables, over which the application of standard rules of statistical physics leads to a dynamics described in terms of Langevin equations. The basic constituents of the model are the following: (i) the movement of ions can be modeled with Langevin equations; (ii) the gate-ions interaction appears as the form of a barrier potential, whose height depends on the state (open or close) of the gate; (iii) the dynamics of the gate is modeled by a variable following a Langevin equation with a potential energy that depends on the membrane potential; and finally (iv) the membrane potential obeys the physics of a capacitor. 

The mechanical variables of the model are the position of the ions $x_i$ (inside the channel domain $[0,L]$) and the gate variables $Y_j$, where $j$ indicates the possibility of several gates. All the physical information concerning these variables are incorporated into a single potential:
\begin{eqnarray}
U(x_i, Y_j, \Delta V) = 
\sum_i V_i(x_i, \Delta V) + 
\nonumber\\
\sum_j V(Y_j, \Delta V) +  \sum_{i,j} V_I(Y_j,x_i).
 \label{pot}
\end{eqnarray}
Here the term $V_i(x_i, \Delta V)$ (see Eq.~\ref{potmem})  is the potential energy originated by the membrane potential $\Delta V$ on the ions inside the pore at position $x_i$. This term is responsible for the physics contained in the GHK equation. The term $V(Y_j, \Delta V)$ (see Eq.~\ref{potgate}) is the potential due to the membrane potential for the dynamics of the $j-$gate, represented by the variable $Y_j$. The last term $V_I(Y_j,x_i)$ (see Eq.~\ref{potint}) corresponds to the interaction of $i-$ions with the $j-$gate. See explicit expressions of these terms in the Appendix. Note that we have neglected ion-ion interactions inside the pore, according with the hypotheses behind the GHK equation. 
Such interactions could straightforwardly be implemented into the approach if more quantitative results were required.

According to this energy functional, the dynamics for the physical variables $x_i$, $Y_j$ is given by the following set of Langevin equations:
\begin{eqnarray}
\gamma_x {\dot x_i} &=& - \partial_{x_i} U(x_i, Y_j, \Delta V) + \xi_i(t),
\label{eqx}\\
 \gamma_{Y_j}{\dot Y_j} &=&  - \partial_{Y_j} U(x_i, Y_j, \Delta V) +  \xi_{Y_j}(t), 
 \label{eqY}
\end{eqnarray}
where thermal noises fulfill
\begin{equation}
\langle \xi_a (t) \xi_b(t') \rangle = 2 \gamma_a \,k_B T\, \delta_{a,b}\, \delta (t- t'),
\end{equation}
and $\gamma_a$ are the corresponding frictions. 

We model thus the motion of the ions inside the channel as a one-dimensional brownian motion, driven by thermal fluctuations, electrostatic potential and the different ionic concentrations  between the two sides of the membrane. Whereas fluctuations and potentials are explicitly put into the Langevin equations, the ionic concentrations appear as boundary conditions at both ends of the channel. 
Regarding gating, we have assumed that the degree of freedom $Y_j$ behaves as a nonlinear spring with two steady states: $Y_C \sim 0$ (close) and $Y_O \sim 1$ (open). This hypothesis is 
similar to the modeling of the gating currents of Ref. \cite{sigg2003fast}, where gating experiments were correlated to a model in which a gating variable undergoes Brownian motion in a one-dimensional diffusion landscape. 

These equations are complemented with the capacitor equation for the dynamics of the membrane potential
\begin{equation}
 C_M \frac{d \Delta V}{d t} = \sum_i I_i,
 \label{capacitorHH}
\end{equation}
where $C_M$ is the membrane capacity assumed to be constant and the r.h.s term includes all the ionic fluxes either across the pore, membrane leaks  or from an external perturbating flux.

The most important feature of this approach is that all interactions between ions and  gates come from the  energy functional of Eq.~\ref{pot}, and hence statistical mechanics can be applied consistently. In particular the dynamical equations verify fluctuation-dissipation relations and  the parameters of the model have a clear physical meaning. Statistical physics consistency is most relevant, since the main features of the results will not depend on parameter fits. The system is autonomous, and the only source of energy, apart from an applied electrostatic potential, is the chemical energy associated with the different ionic concentration  at both sides of the membrane.

Furthermore the degrees of freedom of the gates are coupled to those of the brownian motion of ions by means of the interaction potential $V_I(Y_j,x_i)$. A similar interaction was also considered in Ref. \cite{coalson2008discrete} in the study of a Cl channel. By means of this term the ions see the gate as a barrier potential. As we will see the fact that the model is formulated by a single energy functional implies that 
interaction between ion and gate is mutual.
In particular, gate dynamics could be affected by the collision with ions, which will be of capital importance in the results.



\subsection{A and B pore models}

\begin{figure}[t!]
\centering{\includegraphics[width=0.4\textwidth]{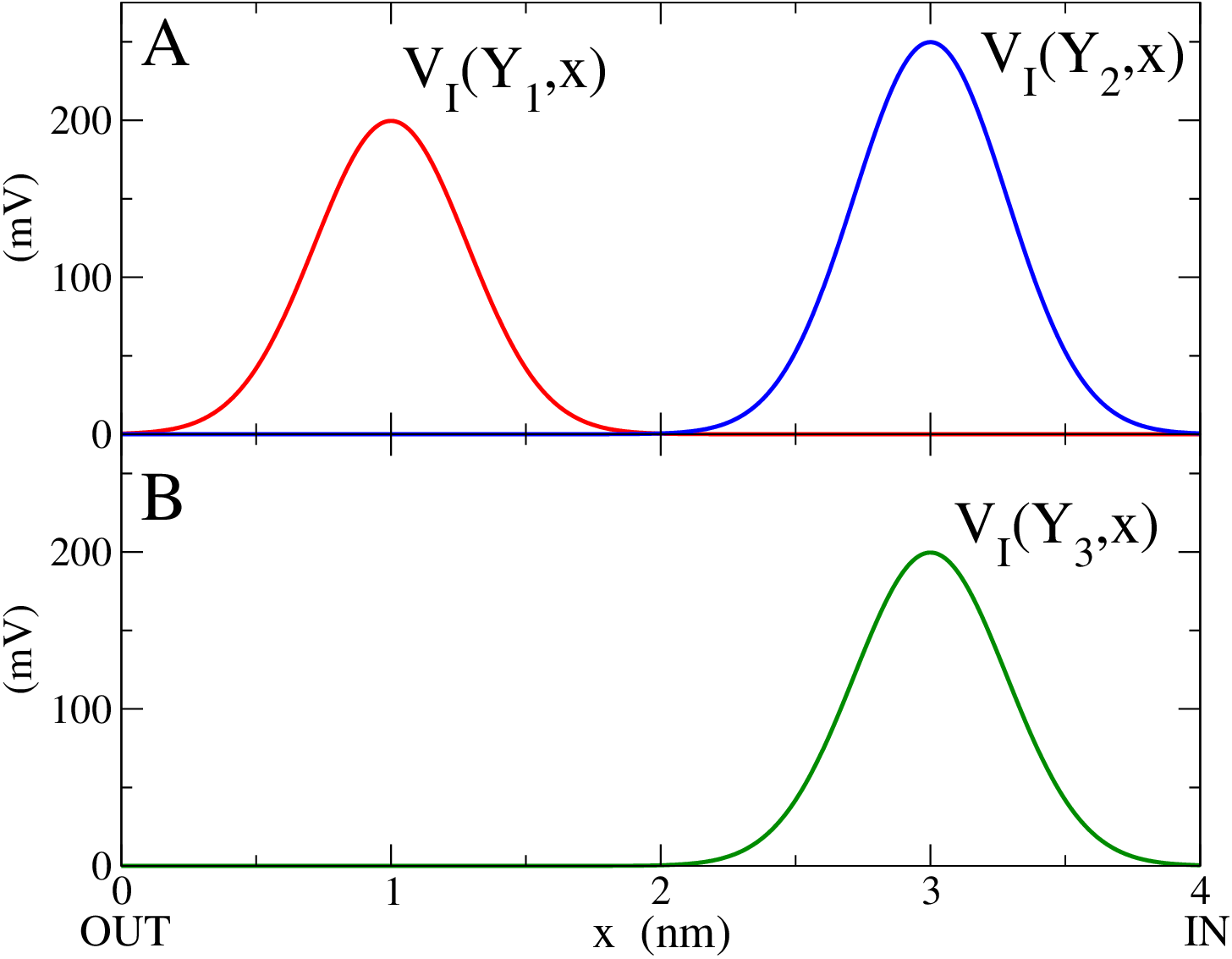}}
\caption{Gate potentials for ions corresponding to both channel models A and B, evaluated for the steady values of closed gates $Y_1=0.029$, $Y_2=0.022$ and $Y_3=0.029$.
}
\label{fig-gates}
\vskip10mm
\end{figure}

We will treat two pore models, A and B, representative of two generic channel families 
for the Na and K ions. 
Regarding the Na permeability, and following notation of Refs.~\cite{H-H,Izhikevich}, the H--H formulation considers two different modulated functions $m(\Delta V)$ (activating) and $h(\Delta V)$ (deactivating).
Accordingly A-type pore will have an activation gate $Y_1$ and an inactivation gate  $Y_2$.  B-type pore will have only one activation gate $Y_3$, corresponding to the modulating function $n$ as the K channel in the literature  \cite{Izhikevich}. Results for these two pores will be compared to available experimental data on Na and K channels.
This theoretical scenario is complemented with the table of parameter values used  to work  in the appropriate biological scale (Table \ref{tableCl}). Graphical representation of the gate effective potentials for both  models are shown in Fig.~\ref{fig-gates}.

\begin{table}[htb]
\begin{center}
\begin{tabular}[c]{|l|l|}

\hline
 $\gamma_{A}$ particle friction & $2 $ $\mu$s meV/nm$^2$ \\
 $\gamma_{B}$ particle friction & $8 $ $\mu$s meV/nm$^2$ \\
$K_B T$ &  $25$ meV  \\
$L$ channel length & $4\, $nm\\
$A$ channel section & $\, $4 nm$^2$\\
$c_0^{\text{A}}(in), c_1^{\text{A}}(out)$ & $0.092, 0.5$ M\\
$c_0^{\text{B}}(in), c_1^{\text{B}}(out)$
& $0.54, 0.075 $ M\\
$C_M$  effective capacity & $1.25$ charges/mV \\

\hline
\end{tabular}
\end{center}
\caption{Physical parameter values used  in the simulations. Both ions have a positive charge $q=+1$ e.}
\label{tableCl}
\end{table}

\section{Results and Discussion}

\subsection{Pore Gating}

\begin{figure}[t!]
\centering{\includegraphics[width=20pc]{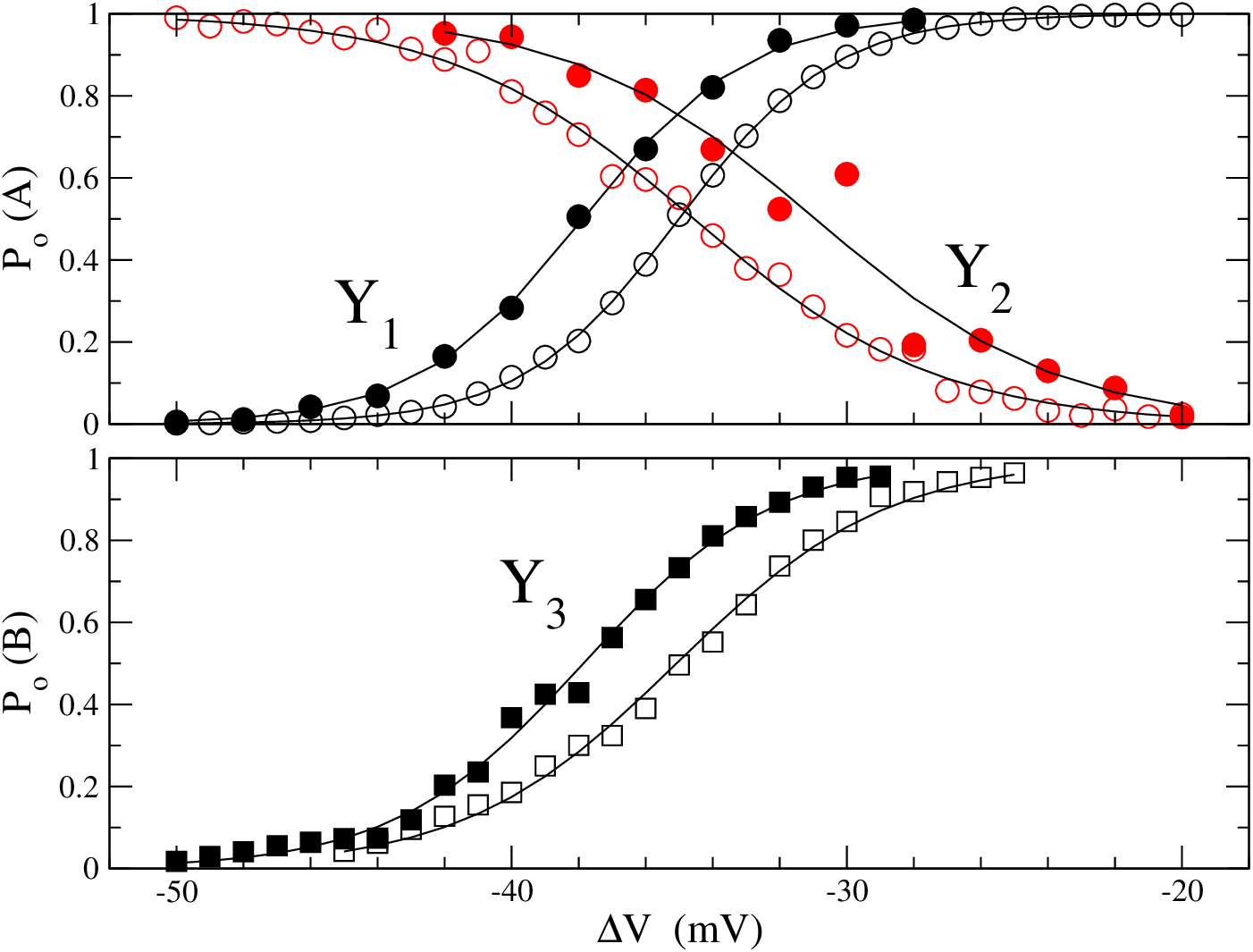}}
\caption{Voltage gating characterization of A and B model  gates. Top: Pore A, $Y_1$ (black) and $Y_2$ (red) gates.
Bottom: Pore B, $Y_3$ gate. Empty symbols: simulations of only gates, {\it i.e.} without ions. Full symbols: simulations with ion concentrations as in Table \ref{tableCl}. Lines are  the fits of expression \ref{P} with the parameters values of tables \ref{tableCl},  \ref{gateparameters2} and \ref{gateparameters}.
}
\label{gating}
\end{figure}

A relevant experimental information on the channel gating is the probability of being open as a function of the membrane potential, $P_o(\Delta V)$ ~\cite{Bezanilla}. We will evaluate this probability for both A and B  models by performing simulations of ion fluxes and gating dynamics by using the Langevin equations \ref{eqx}, \ref{eqY} for ions and gates at fixed values of the membrane potential, as it is done  experimentally. The gate parameters have been chosen to exhibit the experimental gating properties: flux, gating dynamics, time scales, etc (see Tables \ref{tableCl}-\ref{gateparameters2}).

\begin{table}[htb]
\begin{center}
\begin{tabular}[c]{|l|l|l|l|l|l|l|l|l|}

\hline
  & $\gamma$ &  $V_0$& $V_d$& $Q$& $\phi_{ref}$ & a & b &  $x_c$   \\
  &  & \small{k$_B$T} &  \small{k$_B$T}&\small{e} & \small{mV} &    & & \small{nm}  \\
  
\hline
 $Y_1$ & 1000 & 7& 8 & +12 & -35 & 0.2 &  7& 1.0    \\
$Y_2$ & 4000 &  7&10 & -8  & -35 & 0.2 &  9& 3.0  \\
$Y_3$  & 4000 &  7&8 & +10 & -35 & 0.2 & 7 & 3.0  \\

\hline
\end{tabular}
\end{center}
\caption{Parameters of the $Y_1, Y_2, Y_3$ pore gates used in simulations.
Units for $\gamma$ are $\mu$s meV/nm$^2$ and $\sigma= 0.283$ nm.}
\label{gateparameters2}
\end{table}

{\bf Pore A: $Y_1$ and $Y_2$  gates}. 
We proceed with the study of  the two gates of the A model.

{\bf Gate $Y_1$}.  The two equations \ref{eqx} and \ref{eqY} for $j=1$   introduced in the previous section are simulated.
 The second gate $Y_2$ of the channel is also simulated and let fluctuate in its open state, but it is forced to never close. For each value of the membrane potential we record a very long run of a stochastic trajectory, which exhibits roughly rectangular random pulses corresponding to rapid switchings between the two states $Y_1\sim 0,1$. The time spent in the open state, $t_0$, divided by the total time $t_{t}$ gives the open probability for this  potential, $P_o(\Delta V)= t_0/t_t$. 
In Fig. \ref{gating}-top (black dots) we see the results. The empty dots correspond to the case without ions (zero concentrations) and the full dots to the presence of concentrations.  

{\bf Gate $Y_2$}. 
This gate is simulated by the same procedure, but this time it is the $Y_1$ gate which is forced to remain in its open state. Numerical simulation results are plotted in Fig. \ref{gating}-top (red dots). Empty dots correspond to the gate with no ions and the full dots with Na concentrations.
The $P_o(\Delta V)$ of this gate has the same qualitative S--like plot than  that of $Y_1$  but with the steady states interchanged because effective charge has a different sign. 

\begin{figure}[t!]
\centering{\includegraphics[width=0.8\columnwidth]{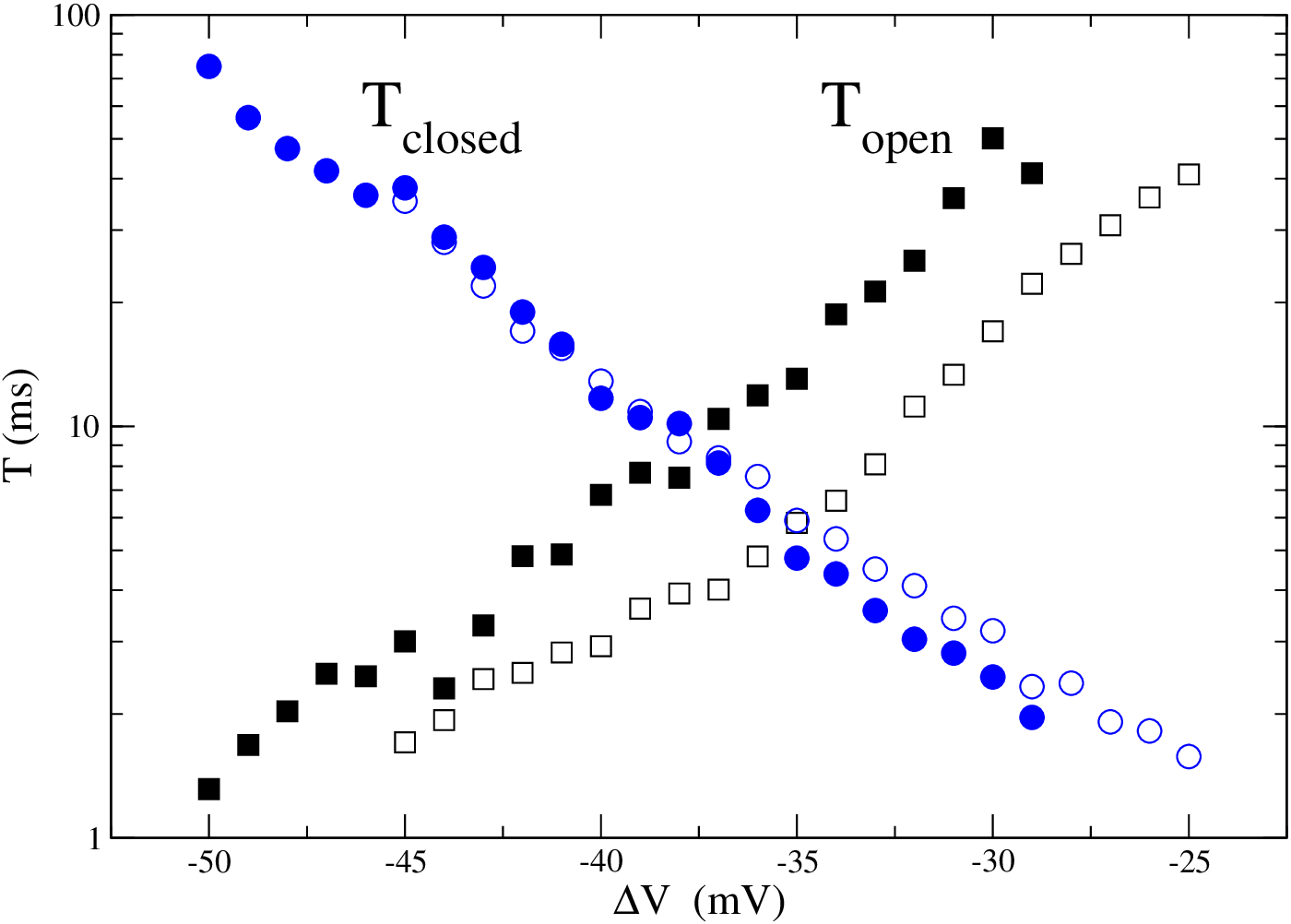}}
\caption{Dynamic characterization of gating for the B model. Blue circles: mean time in the closed state. Black squares: mean time in the open state. Empty symbols: simulations of only gates, {\it i.e.} without ions. Full symbols: simulations with ion concentrations as in Table \ref{tableCl}.
}
\label{times-y3}
\end{figure}

\begin{figure}[t!]
\centering{\includegraphics[width=20pc]{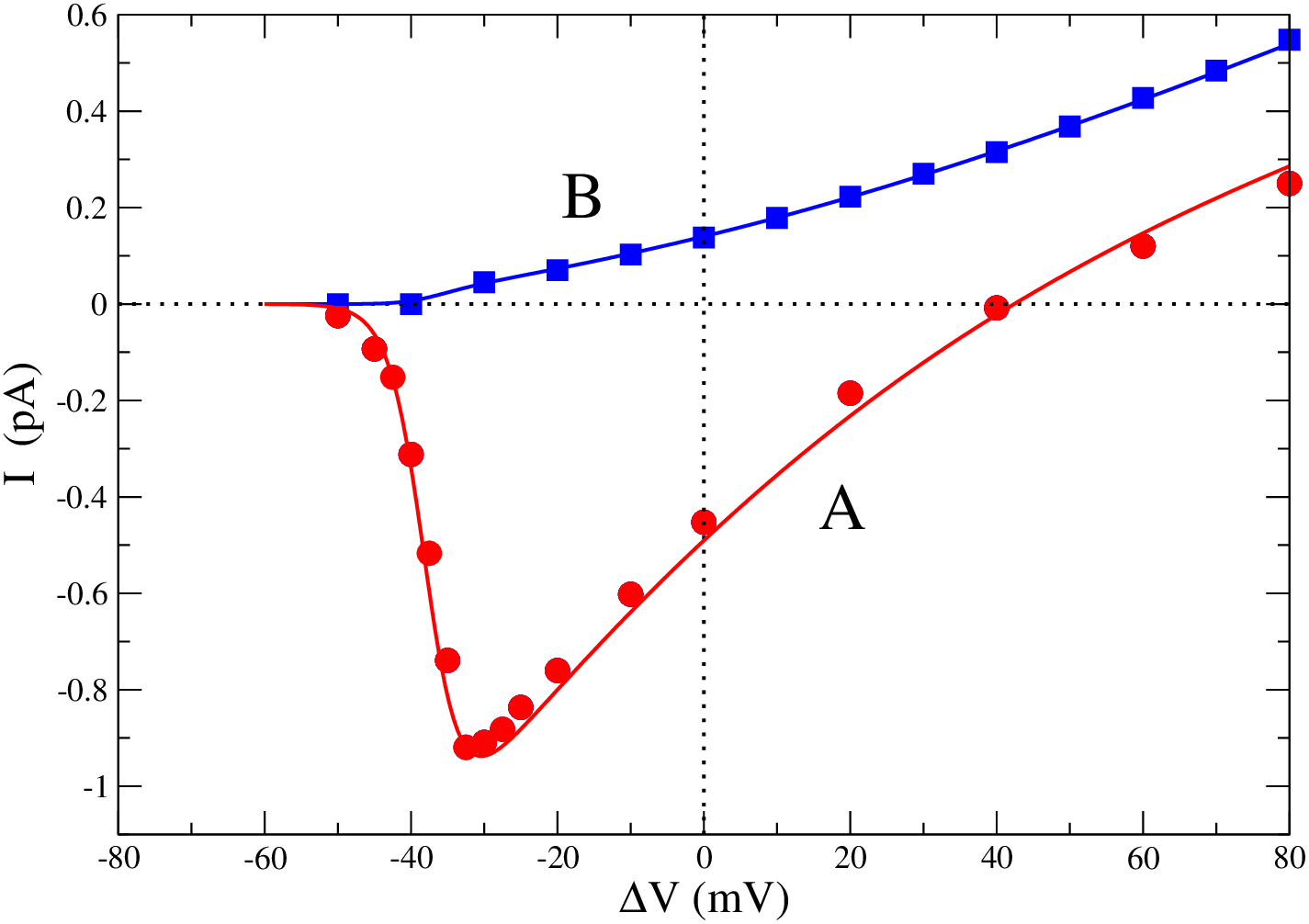}}
\caption{Current peak for A pore (red) and mean current for B pore (blue)   versus the applied voltage $\Delta V$. Lines correspond to the analytical expression \ref{meanI}.
}
\label{gating-flux}
\vskip10mm
\end{figure}

{\bf Pore B: $Y_3$ gate}. 
We proceed with the model B  following the same procedures as in the previous case. This time $Y_3$ is the only gate present in the pore. 
Numerical results are seen in Fig. \ref{gating}-bottom (black squares). Empty symbols correspond to the simulations without ions and full symbols to simulations in the presence of ionic concentrations. We see that $Y_3$ behaves as $Y_1$ but with different parameters. As in the former pore we observe important differences due to the presence of ionic concentrations, and the direction of this effect is the same as in the $Y_1$ case.

We then see in all gates a clear effect of the presence of ions to favor the open state. This result is a consequence of the physical consistency of the approach, in particular of the fact of using a single energy functional for the mutual interaction between ions and gates. 
Gates exert forces on the ions which  implies that they will also have an effect on the gates. 
 As a result the S--like $P_0$ plot moves to smaller potentials for activation gates, and to larger potentials for inactivation gates.

Moreover, in the case of model B we observe that this effect mainly acts by making the closing slower. This can be seen in Fig.~\ref{times-y3}, where the average of the staying times in each state are shown in the same cases as in Fig.~\ref{gating}b. We see that the effect of ions on the closed gate is rather small, since there is only a marginal reduction of the stay time for the smaller polarizations. However there is an important increasing of the stay times for the open state with concentration. In other words the channel closes more slowly in the presence of ions. This effect, well known experimentally \cite{swenson}, is often attributed to a 'foot-in-the-door' mechanism. Note that no specific mechanism has been added to the model to provide such effect. 
On the contrary it has naturally emerged from the physical approach used here.

We return now to the open probability obtained in Fig.~\ref{gating}. The function $P_o(\Delta V)$ is known from well controlled experiments on both single and ensemble of channels. 
The standard explanation for the shape of this function \cite{Hille,phillips,Bezanilla} is that a voltage gating gate has two steady states, open and close, which have different energies
 $U_O(\Delta V)$ and $U_C(\Delta V)$ respectively. Then the characteristic times are weighted by the corresponding Kramers factor for the crossing of a barrier $ t_U \sim \exp{-U/k_BT}$. The relative temporal fraction  of an open state, i.e.  the probability $P_o$ is then,
\begin{eqnarray}
P_o(\Delta V) &=& \frac{e^{-U_O/k_B T}}{e^{-U_O/k_B T} + e^{-U_C/k_B T}} \nonumber \\
&=& \frac{1}{1+e^{-\frac{\Delta U}{k_B T}}} 
 = \frac{1}{2} \left( 1 + \tanh{ \frac{\Delta U}{2 k_B T}} \right),
 \label{P}
\end{eqnarray}
where 
$ \Delta U = U_C - U_O =  Q_{eff} ( \Delta V - \phi_{\text{eff}})
$.

This expression fits very well with the experimental data and accordingly it permits to fix the model internal parameters $Q$ and $\phi_{\text{ref}}$. Nevertheless comparing the values used in our simulations with those obtained by the fitting of the simulation results ($Q_{\text{eff}}$ and $\phi_{\text{eff}}$ in Table \ref{gateparameters}), we observe important differences.
The origin of these discrepancies  is due first  to  the Kramers mathematical approximation and second to the presence of concentrations. 
In our simulations these three gates have the same $\phi_{\text{ref}}= -35 $mV, but different gating charges.  
The fit of Eq.~\ref{P} gives different effective  values for 
$Q_{\text{eff}}$ and $\phi_{\text{eff}}$ depending on whether ions are present ($^c$) or not ($^o$).
It is manifest that the Kramers approximation mainly affects the effective charges and  the concentrations change the effective potential.

\begin{table}[htb]
\begin{center}
\begin{tabular}[c]{|l|l|l|l|l|l|l|}

\hline
  & $Q$ &  $\phi_{\text{ref}}$ & $Q^o_{\text{eff}}$ & $\phi_{\text{eff}}^o$ & $Q^c_{\text{eff}}$& $ \phi_{\text{eff}}^c$ \\
  &e&mV&e&mV&e&mV
  \\
\hline
 $Y_1$ & +12 & -35 & +10.72 &  -35.01& +10.28&-37.86   \\
$Y_2$ & \,\,\,\,-\,\,8 &  -35 & -6.88 & -34.56 &-6.94 &-30.94 \\
$Y_3$  & +10 &  -35 & +7.89 & -35.08 &+8.88 & -37.86\\

\hline
\end{tabular}
\end{center}
\caption{
Gate physical parameters. Second and third columns are the parameter values used  in the simulations, shown for comparison (see Table \ref{gateparameters2}).
The next two columns ($^o$) are their effective values without ion effects, and the last two columns ($^c$) are the effective values when ion concentrations  are present. All these effective values have been obtained from simulation results.}
\label{gateparameters}
\end{table}

%
%

\begin{figure}
\centering{\includegraphics[width=20pc]{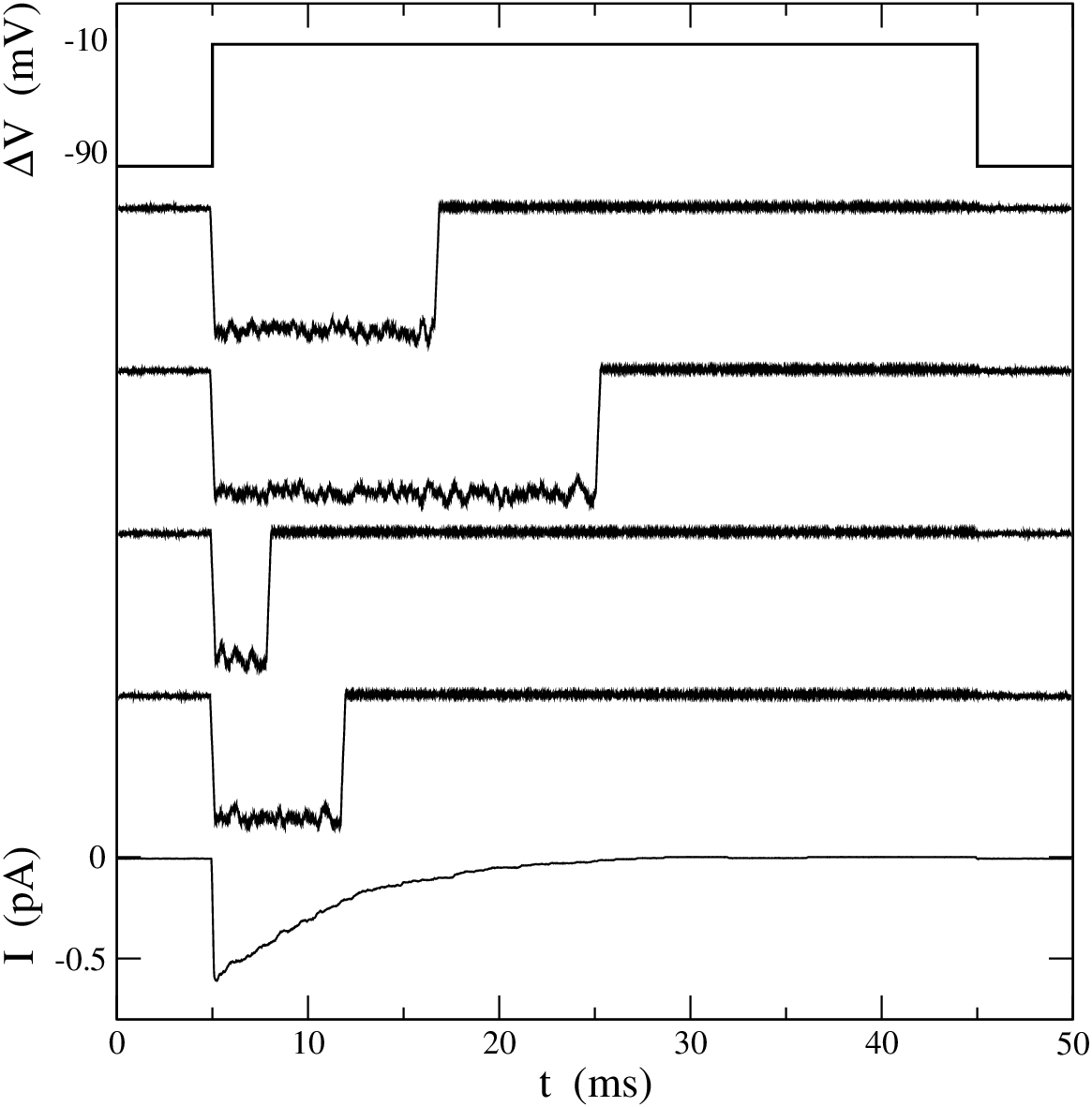}}
\caption{Pulses on the single A model. From top to bottom. A depolarizing step from $-90\,$mV  to  $-10\,$mV  applied to  the potential membrane during $40\,$ms. Fluxes of four pulses under this step, and the  average of 200 events. Intensity signals are filtered by an averaging window of 0.125 ms. 
}
\label{Napulses-1}
\end{figure}

\subsection{Mean ionic fluxes}

By using the effective parameter values of Table \ref{gateparameters} one can calculate the flux of our models from  the  Goldman--Hodgkin--Katz (GHK)  equation  \ref{GHK} for a pore but  modulated by  the expression for $P_o$ in Eq.~\ref{P}, 
\begin{equation}
I(\Delta V)  = P_0(\Delta V) \, \frac{ \Delta V \rho_{in} }{\gamma_x L} \, \frac{ e^ {(V_{Nernst}-\Delta V)/k_B T}- 1}{e^{-\Delta V/k_B T}- 1}\,C_{IJ},
\label{meanI}
\end{equation}
where $C_{IJ} = 0.1602$ pA$\mu$s/n is a conversion factor from particle flux $J$ to intensity $I$.
In Fig.~\ref{gating-flux} we see simulation results (symbols) from the whole model together with theoretical predictions (lines) from Eq.~\ref{meanI}. For the B pore (red symbols) we plot the numerical steady flux,  and for A pores (blue symbols) we represent the current peak. These results are strongly equivalent  to the experiments of  Ref. \cite{Cole} for Na (model A in our model) and K (model B) molecular channels.
One can conclude that GHK equation together with our approach  constitute an appropriate physical scenario to explain experimental results.

\subsection{Dynamics of single pores}

We progress further with the study of the behavior of a single pore, for both A and B cases, under a depolarizing potential step.  Analogous experiments have been key for the understanding of  the internal structure and conformations of the channel, and therefore it is worth to compare the simulations of the model to the available experimental results.

We start with the numerical simulations of the dynamics of a single A pore under the perturbation of depolarizing voltage steps from $-90\,$mV to $-10\,$mV. 
We expect our A model will mimic the experimental results for Na channels ~\cite{Neher,Hammond}.   Four square pulses are seen in Fig. \ref{Napulses-1} during depolarization steps. After the depolarization the gate--1 opens randomly in a short time scale, then  after a larger random time interval gate--2 is closed. Within this interval ions are able to cross the membrane and a larger intensity is observed. The mean average of 200 pulses is a spike--like pulse with fast growing and slow decay. This is the behavior observed in Na single  channel experiments (See Fig. 2 in Ref. \cite{Neher}).

\begin{figure}
\centering{\includegraphics[width=20pc]{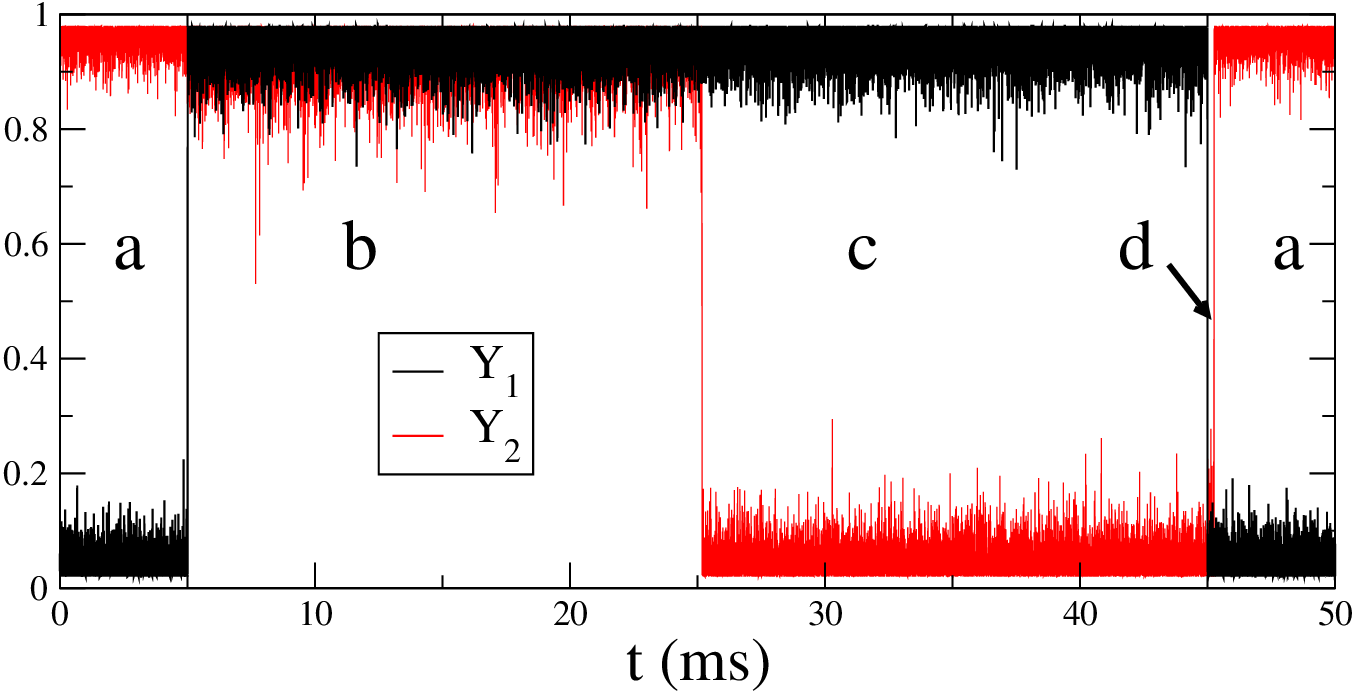}}
\caption{Dynamical evolution of $Y_1$ and $Y_2$ variables (gates)  during the second pulse of Fig.~\ref{Napulses-1}.
}
\label{gates-Na}
\vskip5mm
\end{figure}

In the literature these results are explained by assuming that the Na channel has three possible states: open, close and inactive \cite{Hammond}. In our channel A we have four possible states:
$(Y_1,Y_2) \simeq (0,1), (1,1), (1,0), (0,0)$, shown as $a,b,c,d$ in Fig.~\ref{gates-Na}. These states can be classified as:
close but ready (standby) $a=(0,1)$,  open $b=(1,1)$, and the last two, with $Y_2 \simeq 0$, correspond to inactive refractory states $c=(1,0)$ and $d=(0,0)$. In this figure we see how the duration of the open state, $b$,  is quite random and that the last one, $d$, is very short in time. The  temporal evolution of these variables, going through these states in one depolarizing event is seen in this figure, which corresponds to the second pulse in Fig.~\ref{Napulses-1}.
The three close states could be experimentally discriminated by a finer analysis of the different intensities of the leak flux. This constitutes an additional prediction of the model.

\begin{figure}
\centerline{\includegraphics[width=0.8\columnwidth]{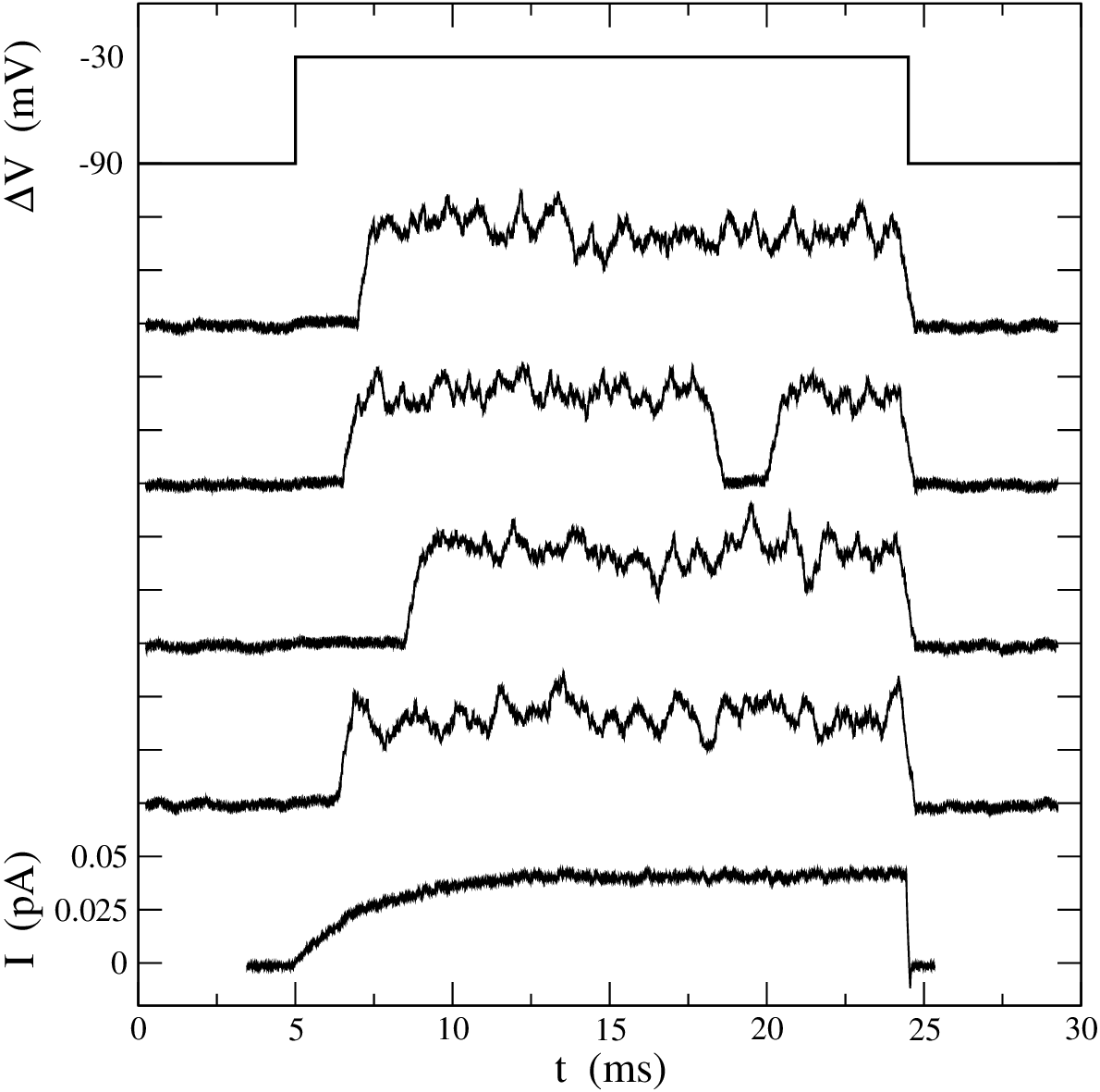}}
\caption{Pulses on the single B channel. From top to bottom. A depolarizing step from $- 90\,$ mV  to  $-30\,$mV  applied to  the potential membrane during $40\,$ms. Fluxes of four  pulses under this step, and the  average over of 200 pulses. Intensity signals are filtered by an averaging window of 0.125 ms.
}
\label{Kpulses}
\end{figure}

The same procedure is next applied to a single B pore with depolarizing voltage steps from  $-90\,$mV  to  $-30\,$mV, as  it is seen in Fig. \ref{Kpulses}. This last voltage corresponds to a high probability for the open state. Numerical results show that the  gate opens quite randomly but it remains open almost all the time until the end of the perturbation. This behavior is similar to that observed in experiments of K single molecular channels  
(see Fig. 4-A from Ref. \cite{Stuhmer}).

\section{Concluding remarks}

We have applied a semi-microscopic Langevin approach to study the gating dynamics of single ionic channels modeled as active pores. { The approach is characterized by the use of a few simple physical mechanisms, variables, and physical laws in a consistent way. This is enough to explain the experimental information concerning gating dynamics, ion fluxes, membrane potential and ionic concentrations. Thus we do not need to consider further biological complexities that would be necessary for more specific observations, such as for example ionic selectivity. }

It has been  assumed here that  ions are charged Brownian particles which follows Langevin equations following standard statistical physics. These pores have gates exhibiting two steady states (open and close) whose dynamics are controlled by nonlinear elastic potentials and Langevin equations.
Ions and gates are treated as mechanical objects interchanging
energy and momentum by mutual collisions, which is accomplished in the model by using a single energy functional.
{ The number of degrees of freedom used is thus maintained to a minimum. This permits to isolate the relevant mechanisms for the studied phenomenology and to obtain results comparable to experiments without the need of additional detailed channel structure. Moreover
the physical nature of the approach 
permits, in a straightforward way, its extension for a more quantitatively detailed study and for taking into account additional mechanisms in other channels or alternative gating modelings.}

We have presented two pore models representing Na and K molecular channels.
Their gating dynamics, as represented by the open probability of the pores and by the steady  and peak ionic currents as functions of membrane potential, exhibit the main characteristics observed in experiments on single channels \cite{Hille,Hammond}. 

Additionally, we have obtained the effect that concentration has on channel gating. Namely, the increasing of ion concentration enhances the probability of the open state. Thus the open-probability curves move to smaller values of the membrane potential for activation gates, and to higher values for inactivation ones. 
Very similar effects, associated to the so called ``foot-in-the-door mechanism'', have extensively been studied experimentally, mainly in K channels \cite{swenson} but also in other channels \cite{cheng2007modeling,teijido2014}. 
{ This is a most relevant result of our model, and it is particularly interesting the fact that }
it is a direct consequence of having a single energy functional for the mutual interaction between gates and ions, which constitutes a necessity of physical consistency.
Therefore this effect should have a general validity for permeant ions, not depending on the specific structure of the channel.

This approach opens interesting perspectives. Since all the model parts are described by standard and well controlled  physical laws one can address particular aspects for single channels or single gating events.
Moreover
model A implies the existence of three closed gate states for the Na channel, which could be discriminated by measuring leak charge intensities through the channel. This prediction calls for experimental verification.

Channel parameters are obtained  from experimental data and can be different for channels of the same family.  These differences can enlighten a variety of  internal channel structures which can help to refine molecular descriptions. In other situations different mechanical types are possible. 
Also,  
we have shown the specific effect of the ionic cell concentrations on the gating process of each gate. 
Our results show the interest in exploring further this effect.
%

Further numerical refining could consist of including ion-ion interactions, which were omitted for simplicity, and due to the fact that it was not necessary to reproduce the basic phenomenology. Also several additional activation gate variables could be implemented for each channel, to account for the tetrameric structure  of  four voltage sensing domains in the Na and K channels \cite{Hammond}. 
These and other possible details, such as more complex potential landscapes for both ions and gates \cite{sigg2003fast}, are straightforward in this framework and could be important in order to use it for a more quantitative modeling.

Finally our approach can be  extended to other possible channel configurations, such as to other channels from the Na$^+$ and K$^+$ families,   and also  to model Ca$^{2+}$ or Cl$^-$  channels. 
Also, by using this approach, it would be interesting to introduce ionic diffusion out of the membrane, in order to study the coupling among regions of the membrane \cite{jianxue1997} or with cell vesicles \cite{dawson2002}.

\vskip3mm

 {\bf Acknowledgements:}This work was supported by the Ministerio de Economia y Competitividad (Spain) and FEDER (European Union), under
 projects FIS2015-66503-C3-2P/3P
 and by the Generalitat de Catalunya Projects 2014SGR1093 and 2014SGR878.


\appendix

\setcounter{figure}{0} 
\renewcommand{\theequation}{A-\arabic{equation}}
\renewcommand{\thefigure}{A-\arabic{figure}}
\renewcommand\thesection{Appendix \Alph{section}:}

\section{Semi-microscopic approach}

\renewcommand\thesection{\Alph{section}}

\subsection{The electrostatic potential $V_i(x_i, \Delta V)$}

 The theoretical framework for the ion dynamics is the known Goldman--Hodgkin--Katz (GHK)  equation \cite{Goldman}, for the flux through a pore, that we summarize here.
Free ions inside the pore are described by point--like particles of charge $q$ moving in a one dimensional space at position $x_i(t)$ under the effective electrostatic membrane potential 
\begin{equation}
V_i(x_i, \Delta V)= \frac{q\Delta V}{L} (x_i- L), \qquad 0 < x_i < L,
\label{potmem}
\end{equation}
where $\Delta V$ is the potential difference between both sides of the membrane, and L is the length of the pore.

\subsection{The gate potential $V_j(Y_j,\Delta V)$}

 The evolution of this variable is controlled by the  nonlinear elastic  potential,
\begin{eqnarray}
 V(Y,\Delta V)&=& V_0 \left[ -a \ln (Y(1-Y)) - b (Y-0.5)^2 \right] 
 \nonumber\\
 &+& Q(\Delta V - \phi_{\text{ref}}) Y.
 \label{potgate}
 \end{eqnarray}
The first part  has the form of a double well potential, and  refers to the internal structure of the pore responsible for the gate bistability. 
The parameter values $V_0, a, b$ have to be chosen to enter into the experimental scale. 
Note that by choosing $a \ll b$ the well minima are very close to $0,1$, specifically at $Y\simeq\frac{a}{b},1-\frac{a}{b}$. A similar approach was used in \cite{cheng2007modeling,parc2009} for the gate in a Cl channel.
Other expressions for the potential can be used, since
the specific form is not really important, provided it has two minima, corresponding to open and close states, separated by an energy barrier. Other forms for this double well potential have been explored, giving essentially similar results. Note that the only relevant physical parameter of the bistable potential is the height of the barrier.

The last term is the interaction with the membrane potential where $Q$ is the  charge of the gate sensor and  $\phi_{\text{ref}}$ is the reference potential that determines the $\Delta V$ value at which both states are equally probable. These last two parameters are characteristic of a specific channel and  their values  have to be obtained from experimental data (see for example experiments in Ref.~\cite{Bezanilla}). 

\subsection{The ion-gate interaction potential $V_I(Y,x_i) $}

The interaction between ions and gate is modeled by a potential energy, which represents a physical barrier for the ions. This barrier has a specified position inside the channel and a prescribed width. Its height is variable and controlled by the state of the gate represented by the $Y$ value.
The proposed potential is 
\begin{equation}
 V_I(Y,x_i) = V_d f(Y)  \exp \left(-\frac{(x_i - x_c)^2}{2 \sigma^2} \right).
 \label{potint}
\end{equation}
Here, $x_i$ is the ion position, $x_c$ is the  the center of the gate potential inside the channel,  and $\sigma$ is its width. 
Note that, in view of Eqs.~\ref{eqx},\ref{eqY}, this potential will produce mutual forces on both ions and gate.

The height of the barrier is modulated by the function $f(Y)$, which defines a correspondence between the state of the gate (the $Y$ value) and the height of the barrier  $V_d f(Y)$ seen by the ions. Thus the necessary conditions for the function $f(Y)$ are: $f(0)=1$ for the close gate (maximum barrier height), $f(1)=0$ for the open gate (minimum barrier height). Additionally, since Y is a fluctuating variable, we construct f(Y) as having zero slopes at these values so the barrier height does not present large fluctuations while the gate remains in the same state.
For the modulating function $f(Y)$
the  envelope function $f(Y) = (1 + \cos \pi Y)/2$ is used. Other expressions for this function have been tested, leading to very similar results.

\subsection{Numerical methods}

Our whole system is composed of three physical domains or volumes: the channel, along which the ions move and where the gates are placed, and two reservoirs at both channel ends, corresponding to the regions inside and outside the cell respectively. 

The dynamics inside the channel are driven by Langevin equations \ref{eqx} and \ref{eqY}. The numerical integration of these equations 
is performed with a standard explicit (first order Euler) algorithm.
The employed time step has been $\Delta t = 1.25\times 10^{-4}$ $\mu$s for the complete system (gates and ions) and $\Delta t=10^{-2}$ $\mu$s when we had only gates (zero external ion concentrations).  In particular ions move through a pore of length $L$, and they disappear from simulation whenever the position $x_i$ escapes from the interval $(0,L)$.

The reservoirs are implemented as boundary conditions at both ends of the pore for the Langevin dynamics of ions. That means that the ionic concentration values at the reservoirs determine the rate at which ions enter into the pore.

The numerical simulation of these equations allows to record the evolution of  the state of the gates and the number of particles crossing the boundaries. In those cases where the membrane potential is not externally fixed the membrane potential $\Delta V$ is updated by using the numerical integration of the capacitor equation (Eq.~\ref{capacitorHH}) in the form
\begin{equation}
 \Delta V (t + \Delta t) = \Delta V (t) - \frac{\Delta Q_{1} + \Delta Q_{2} + \Delta Q_\text{ext}}{2 C_M},
 \label{capacitor-discret}
\end{equation}
where $\Delta Q_1$ and $\Delta Q_2$ are  the total charge crossing each of the two boundaries in the time interval $\Delta t $. The additional term $\Delta Q_\text{ext}$ accounts for external inputs to the membrane charge. 
The divisor ``2'' takes into account that charges which cross both
 boundaries  are being counted twice.
The final output are the values of the membrane potential $\Delta V$ and of the ionic fluxes $J = \Delta Q/\Delta t$,  which will be compared with  known experimental results.

\subsection{Validation of the stochastic approach}

A necessary test of the model is the dynamical relaxation of the open channel to its steady state  in the presence of fixed ion concentrations at the boundaries. This is done for both models.
The evolution of the membrane potential to the Nernst value corresponding to the concentrations at both sides of the membrane (without any parameter fitting) supports the consistence of the dynamical model, 
and also the correctness of the used algorithms. 


From the theoretical point of view ions obey the  overdamped Langevin equation,
\begin{equation}
 \gamma_x {\dot x_i} = - \partial_{x_i} V_i(x_i) + \xi_i(t),
\label{eqx1}
\end{equation}
where $\gamma_x$ is the effective friction and $\xi_i(t)$ is a thermal noise of zero mean and correlation $\langle \xi_i(t)\,\xi_i(t') \rangle = 2 \gamma_x k_B T \delta(t-t') $. 
One can consider here the Fokker--Plank equation for the density of ions inside the channel. 
Assuming steady state with constant flux, $J(x,t)=J$, and the 
boundary conditions for the one-dimensional ionic densities at both sides, $\rho_{in}$ and $\rho_{out}$, we get the well known Goldman--Hodgkin--Katz equation,
\begin{equation}
J (\Delta V)=  \frac{q\Delta V \rho_{in}}{\gamma_x L} \frac{e^{q(V_\text{Nernst}-\Delta V)/k_B T}- 1}{ e^{-q\Delta V/k_B T}-1},
\label{GHK}
\end{equation}
where $V_\text{Nernst}= ({k_B T}/{q}) \ln{c_{in}/c_{out}}$ is the Nernst potential corresponding to these concentrations, and $q$ is the ion charge. Note that the one-dimensional densities $\rho_{in/out}$ are related to the bulk concentrations $c_{in/out}$ as  $\rho_{in/out}= A\;c_{in/out}$, where $A$ is the effective section of the channel.
This equation is a relevant analytical reference when the channel is in the open state.

\begin{figure}[t!]
\centering{\includegraphics[width=20pc]{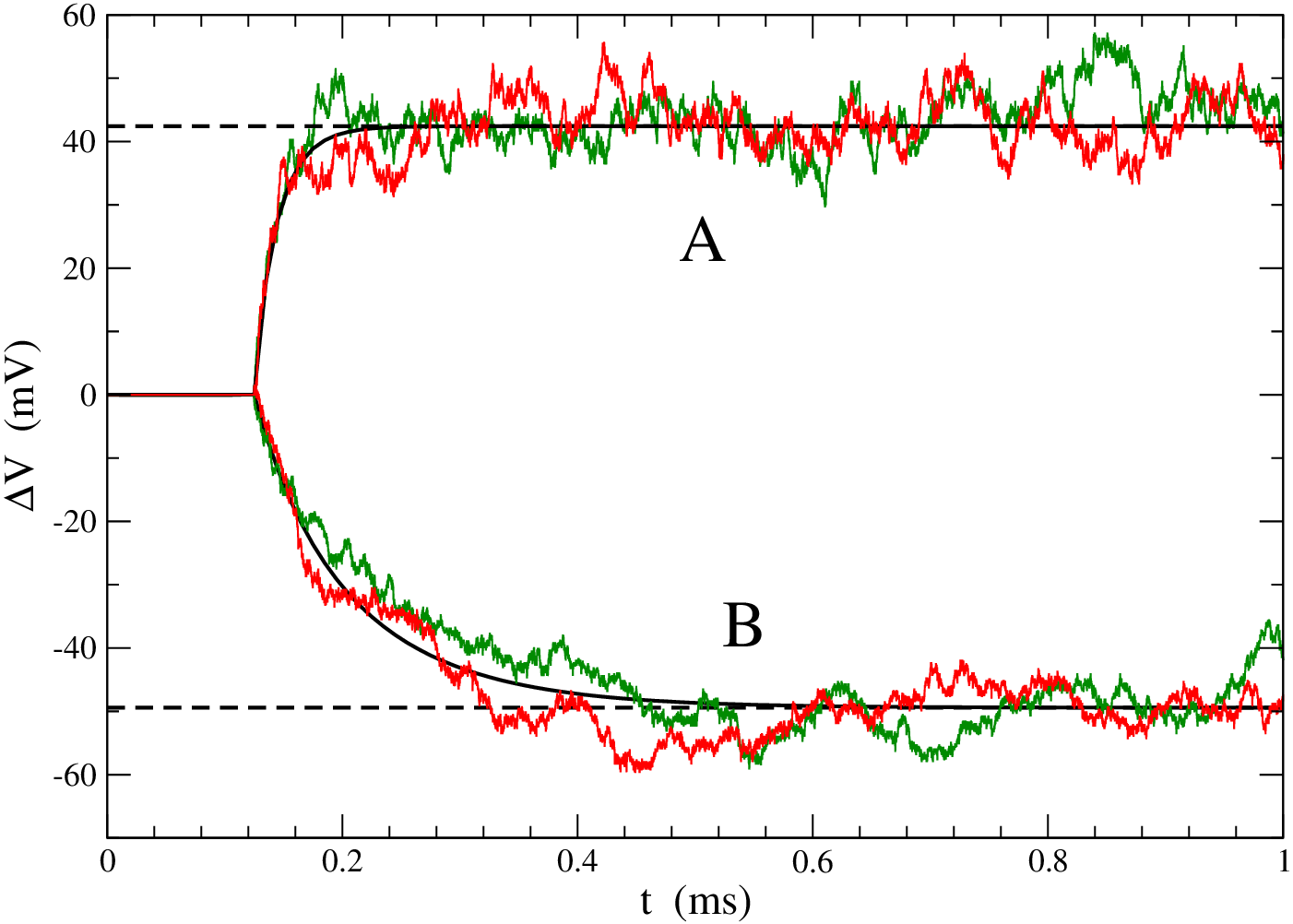}}
\caption{Time evolutions (colored curves) of the membrane potential induced by  the ions of models A  (upper curves) and B  (lower curves) obtained from stochastic simulations. The potential membrane evolves to the corresponding Nernst potentials  $V_{A} \simeq 42.41\,$mV   and $ V_B \simeq -49.43\,$mV. Full black lines correspond to the numerical integration of Eq. \ref{Nernst} and broken lines indicate the asymptotic Nernst values (See text).
}
\label{channel-1K}
\end{figure}

In the simulations we initially populate the pore letting ions evolve with a fixed $\Delta V = 0$ (this would correspond in experiments to electrically connecting both membrane sides). At  $t= 125$ $\mu$s the voltage is then left free to evolve.
From the stochastic evolution of the charges  we record the balance $\Delta Q$ crossing each of the channel boundaries  during the time step $\Delta t$. That includes particles hopping out of the system and particles entering into it through that boundary. Then from Eq.~\ref{capacitorHH} the change in the membrane potential is evaluated and recorded. 
Two realizations of each channel model are shown in Fig. \ref{channel-1K}. 

This simulation result is complemented with the theoretical calculation of the membrane potential  from capacitor equation \ref{capacitorHH} and the GHK equation \ref{GHK}, which results in the differential equation
\begin{equation}
 \frac{d\,\Delta V}{ d t} = 
 -\frac{J(\Delta V) }{C_ {eff} }
 \label{Nernst}
\end{equation}
The numerical integration of this equation provides a prediction for the deterministic evolution of $\Delta V(t)$. The Nernst potential is the final steady value. This results should be compared to the stochastic evolution obtained from the numerical simulation of the Langevin equations. This can be seen in Fig. \ref{channel-1K}, where two stochastic evolutions of $\Delta V(t)$ are shown for each model (A and B). It is worth to remark that they are single trajectories without any statistical average. These evolutions behave as expected for both channels.  In view of Eq.~\ref{Nernst} the difference in the times scales for both ions correspond to the difference in the friction coefficients $\gamma_x$ and in the ion concentrations present in each channel.

\bibliographystyle{epj}
\bibliography{NaK}

\end{document}